\documentstyle[preprint,aps,psfig]{revtex}

\newcommand{\be}{\begin{eqnarray}}
\newcommand{\ee}{\end{eqnarray}}
\newcommand{\bc}{\begin{center}}
\newcommand{\ec}{\end{center}}
\newcommand{\pa}{\partial}
\newcommand{\fr}{\frac}

\begin{document}
\draft
\tighten

\title { A TRANSMISS\~AO DO CALOR EM RELATIVIDADE}
\author{ A. Brotas\footnote{E-mail:brotas@fisica.ist.utl.pt} ,
    J.C. Fernandes\footnote{E-mail:joao.carlos@tagus.ist.utl.pt},   }
\address{ Departamento de F\'{\i}sica, Instituto Superior T\'ecnico, \\
Av Rovisco Pais 1096.  Lisboa Codex, Portugal}

\date{\today}

\maketitle

\begin{abstract}

The simultaneous study of deformation and heat transmission in a bar was
ignored for about 150 years.

The traditional Fourier equation just allows to study the evolution of
temperature in a undeformable bar . The search for its relativistic
variant is a task which must fail because in Relativity there are no
undeformable bars. Rigid bodies,in the sense as rigid possible, are
deformables.

In this work we show how to write in Relativity the system of equations
necessary to study simultaneously deformation and temperature evolution
along a rigid bar. \\  \\  \\

\bc {\begin{large}  Resumo \end{large} } \ec

	O estudo conjunto da deforma\c c\~ao e da transmiss\~ao do calor numa barra 
 foi ignorado durante quase 150 anos. 

	A tradicional equa\c c\~ao de Fourier s\'o permite estudar a 
evolu\c c\~ao da temperatura ao longo de uma barra indeform\'avel. 
A procura de uma sua variante relativista 
\'e uma tarefa votada ao insucesso porque em Relatividade n\~ao h\'a 
barras indeform\'aveis (os corpos r\'{\i}gidos no sentido o mais r\'{\i}gidos 
poss\'{\i}veis s\~ao deform\'aveis).

	Neste texto mostramos como escrever, em Relatividade, o sistema de 
equa\c c\~oes necess\'ario para estudar em conjunto a deforma\c c\~ao e a 
evolu\c c\~ao da temperatura ao longo de uma barra r\'{\i}gida.

\end{abstract}
\newpage


\section{Uma tentativa sem futuro}

	A procura de uma variante da equa\c c\~ao de Fourier que exclua a
possibilidade de transmiss\~ao de energia e de sinais a uma velocidade 
superior a $c$, que tem ocupado alguns f\'{\i}sicos desde o in\'{\i}cio da 
Relatividade, \'e uma tarefa votada ao insucesso pela raz\~ao seguinte: \\

	Em F\'{\i}sica Cl\'assica admitimos que o corpo r\'{\i}gido (no 
sentido o mais r\'{\i}gido poss\'{\i}vel) \'e indeform\'avel. A equa\c c\~ao 
de Fourier, equa\c c\~ao da transmiss\~ao do calor numa barra 
indeform\'avel \'e, assim, a equa\c c\~ao do fen\'omeno f\'{\i}sico: 
transmiss\~ao do calor numa barra r\'{\i}gida. \\ 

	Em Relatividade, r\'{\i}gido e indeform\'avel n\~ao s\~ao sin\'onimos
(ver, por exemplo, \cite{um}). O corpo r\'{\i}gido no sentido o mais 
r\'{\i}gido poss\'{\i}vel \'e deform\'avel. Uma variante da equa\c c\~ao de 
Fourier que estude a transmiss\~ao do calor numa barra indeform\'avel 
n\~ao pode servir, assim, para estudar o  fen\'omeno f\'{\i}sico: 
transmiss\~ao do calor numa barra r\'{\i}gida. Este fen\'omeno s\'o pode ser 
abordado no \^ambito do estudo da transmiss\~ao do calor em corpos 
deform\'aveis. \\ 

	Para precisar ideias, consideremos o caso de duas barras 
semi-infinitas, uma com uma temperatura uniforme $T_1$ e outra com uma 
temperatura uniforme $T_2$ , que s\~ao postas em contacto no instante 
$t = 0$. Em F\'{\i}sica Cl\'assica, a equa\c c\~ao de Fourier permite-nos 
estudar a evolu\c c\~ao da temperatura nas barras {\it no caso de elas serem 
indeform\'aveis (r\'{\i}gidas)}. No caso, por\'em, de serem deform\'aveis, as 
variac\c c\~oes de temperatura e de press\~ao provocam deforma\c c\~oes. 
O problema \'e muito mais complicado e n\~ao pode, em rigor, ser estudado com 
uma \'unica equa\c c\~ao. Precisamos de um sistema de equa\c c\~oes (duas no caso 
a uma dimens\~ao). \\ 

	Embora os dois fen\'omenos da deforma\c c\~ao e da transmiss\~ao 
do calor se verifiquem em simult\^aneo na vulgar vibra\c c\~ao de uma 
mola, a F\'{\i}sica Cl\'assica ignorou o problema da transmiss\~ao 
do calor nos meios deform\'aveis durante quase 150 anos. Por duas 
raz\~oes: porque era um problema dif\'{\i}cil e porque n\~ao surgiu 
nenhum problema, nem pr\'atico, nem te\'orico, que exigisse o seu estudo. \\

	Num outro texto apresentaremos uma abordagem do 
problema em F\'{\i}sica Cl\'assica. Neste, abordamos directamente o 
problema relativista, a uma dimens\~ao, no caso limite dos corpos r\'{\i}gidos.


\section{A vibra\c c\~ao de uma barra el\'astica}

\subsection{Sem transmiss\~ao do calor}

	Comecemos por abordar o problema: como escrever em Relatividade 
a equa\c c\~ao do movimento adiab\'atico, isto \'e, sem transmiss\~ao do 
calor, de uma barra el\'astica? \\ 

	O processo mais elegante para o fazer \'e o seguinte: \\ 

	Sendo $X^i$ as coordenadas "fixas" dos pontos da barra e $x^i,x^4=c ~ t$ 
as coordenadas de um referencial de in\'ercia $S$, o movimento da barra 
pode ser descrito em representa\c c\~ao de Lagrange por: $x=x(X,t)$ , ou em 
 representa\c c\~ao de Euler por: $X=X(x,t)$ . \\ 

	A velocidade de cada ponto $X$ da barra em rela\c c\~ao ao 
referencial de in\'ercia $S$ \'e dada por: 

\be v ~ = ~ \frac{\pa x}{\pa t} = - \frac{\frac{\pa X}{\pa t}}
{\frac{\pa X}{\pa x}} \ee

	O comprimento de cada elemento $dX$ da barra sofre, normalmente, 
deforma\c c\~oes ao longo do tempo. O seu comprimento pr\'oprio num dado 
instante \'e dado por: 

\be dX_p = \fr{\fr{\pa x}{\pa X}dX}{\sqrt{1 - \beta^2}} ~~~~ ; ~~~ com ~~ 
\beta = \fr{v}{c} \ee 

No caso da vari\'avel $X$ ter sido escolhida de modo a que $dX$ seja 
o comprimento do elemento $dX$ da barra quando n\~ao deformada, o estado 
de deforma\c c\~ao da barra \'e caracterizado por um $s$ dado por: 

\be s = \fr{dX_p}{dX} = s(X,t)  = \fr{\fr{\pa x}{\pa X}}{\sqrt{1 - \beta^2}} 
= \fr{1}{\sqrt{\left( \fr{\pa X}{\pa x}\right)^2 - \fr{1}{c^2} 
\left( \fr{\pa X}{\pa t}\right)^2}} \ee

	Num corpo el\'astico, a press\~ao e a densidade no referencial 
pr\'oprio local, que representamos por $p$ e $\rho_0$ , devem ser 
fun\c c\~oes da deforma\c c\~ao $s$ e da temperatura $T$ :

\be \begin{array} {lr} p = p(s,T) ~~~~ & ~~~~ \rho_0 = \rho_0(s,T) 
\end{array} \ee 

	Sobre uma dada adiab\'atica {\small $ad$} devemos ter:

\be T = T_{ad} (s) \ee 

	Assim, no caso de n\~ao haver transmiss\~ao de calor, no estudo de um movimento
 de uma barra, devemos poder escrever: \footnote[1]
{Designando por $\rho_0^0$ a densidade do material n\~ao deformado $(com ~~ s=1)$ no 
referencial pr\'oprio 
a uma dada temperatura $T_0$ (que caracterizar\'a uma dada adiab\'atica {\small $ad$}),
 a conserva\c c\~ao da energia imp\~oe-nos (na compress\~ao 
ou extens\~ao adiab\'atica de um elemento de comprimento $l_0$ da barra): 
 \bc $  - l_0 \int_1^s p_{ad} ds = (\rho_{0ad}.s - \rho_0^0) l_0 c^2 $ \ec 
o que nos permite escrever: 
 \bc $ p_{ad}(1) = 0 ~~ ; ~~ p_{ad} = - c^2(\rho_{0ad} + \fr{d\rho_{0ad}}{ds}.s) $ \ec 
	Todas as f\'ormulas da elasticidade relativista adiab\'atica t\^em de respeitar esta
rela\c c\~ao. \'E o caso das f\'ormulas correspondentes ao caso limite dos materiais 
r\'{\i}gidos usadas em (10) em que escrevemos simplesmente $p$ e $\rho_0$ em vez 
de $p_{ad}$ e $\rho_{0ad}$.} \\ 

\be p = p(s, T_{ad}(s)) = p_{ad}(s) ~~~~ ; ~~~~ \rho_0 = \rho_0(s, T_{ad}(s)) = \rho_{0ad}(s) \ee

	Interessa-nos considerar o referencial de in\'ercia $S^*$ de coordenadas 
$(x^*, x^{4*} = ct^*)$ que, no instante $t$, acompanha cada ponto $X$ da barra, de 
velocidade $v = v(X,t)$. \\

	No instante $t$, as componentes, em $S^*$, do tensor impuls\~ao energia do meio
material da barra na vizinhan\c ca do ponto $X$ s\~ao:

\be T^{\alpha^*\beta^*} = \left[ \begin{array}{cccc} 
p & 0 & 0 & 0 \\
0 & 0 & 0 & 0 \\
0 & 0 & 0 & 0 \\
0 & 0 & 0 & \rho_0 c^2 \end{array} \right] \ee

	As f\'ormulas de transforma\c c\~ao entre $(x,t)$ e $(x^*,t^*)$ s\~ao as 
f\'ormulas de transforma\c c\~ao de Lorentz. \\ 

	As componentes do mesmo tensor $T^{\alpha \beta}$ em $S$ s\~ao, em consequ\^encia: 

\be T^{\alpha \beta} = \left[ \begin{array}{cccc}
\fr{p + \beta^2 \rho_0 c^2}{1 - \beta^2} & 0 & 0 & \fr{\beta(p + \rho_0 c^2)}{1 - \beta^2} \\ 
0 & 0 & 0 & 0 \\ 
0 & 0 & 0 & 0 \\
\fr{\beta(p + \rho_0 c^2)}{1 - \beta^2} & 0 & 0 & \fr{\beta^2p + \rho_0 c^2}{1 - \beta^2} 
\end{array} \right] \ee 

	Podemos fazer o mesmo racioc\'{\i}nio em todos os instantes $t$, para todos os pontos
 $X$ da barra (tendo, naturalmente de considerar referenciais $S^*$ diferentes). \\ 

	Conhecidas as f\'ormulas (6) podemos conhecer as componentes, em 
qualquer referencial de in\'ercia $S$, do tensor impuls\~ao energia em coordenadas 
$(x, x^4=ct)$ de $S$, em todos os pontos $X$ e em todos os instantes $t$. \\ 

	As leis de conserva\c c\~ao exprimem-se, no caso do movimento a uma dimens\~ao, 
pelas duas equa\c c\~oes: 

\be \partial_\alpha T^{1\alpha} = 0 ~~~~ ; ~~~~ \partial_\alpha T^{4\alpha} = 0 \ee 

	Consideremos, a t\'{\i}tulo de exemplo, o movimento de uma barra r\'{\i}gida. 
Neste caso, as leis el\'asticas (adiab\'aticas) s\~ao: 

 \be p = \fr{\rho_0^0 c^2}{2} \left[ \fr{1}{s^2} - 1 \right] ~~ ; ~~ 
\rho_0 = \fr{\rho_0^0}{2} \left[ \fr{1}{s^2} + 1 \right] \ee

	Usando estas leis e desenvolvendo os c\'alculos obtemos as duas equa\c c\~oes: 

\be \begin{array}{c} \fr{\partial^2X}{\partial x^2} - \fr{1}{c^2} \fr{\partial^2X}
{\partial t^2} = 0 \\  \\ 
\fr{\partial X}{\partial t} \left( \fr{\partial^2X}{\partial x^2} - \fr{1}{c^2} 
\fr{\partial^2X}{\partial t^2} \right) = 0 \end{array} \ee

	S\~ao estas as equa\c c\~oes do movimento relativista de uma barra r\'{\i}gida. 
A primeira traduz a conserva\c c\~ao da quantidade de movimento e a segunda a 
conserva\c c\~ao da energia. \\ 

	De facto, obtivemos um sistema de duas equa\c c\~oes, mas como as solu\c c\~oes 
da primeira s\~ao solu\c c\~oes da segunda, na pr\'atica corrente ignoramos a segunda 
equa\c c\~ao e esquecemos o sistema que resultou da aplica\c c\~ao dos dois 
princ\'{\i}pios de conserva\c c\~ao. Ora, \'e este sistema de equa\c c\~oes que nos 
vai permitir abordar, como vamos ver adiante, o problema conjunto do movimento e da 
transmiss\~ao do calor. \footnote[2] 
	{No caso de considerarmos outras leis el\'astica adiab\'aticas, os c\'alculos 
s\~ao bastante mais longos, mas o resultado final \'e o mesmo: chegamos a um sistema 
de duas equa\c c\~oes em que as solu\c c\~oes da primeira s\~ao solu\c c\~oes da segunda.}
\footnote[3] 
{Se conhecermos $T = T(s,p)$, podemos calcular $T = T(X,t)$, mas, nos problemas
sem transmiss\~ao do calor, este resultado 
aparece-nos no final e pode ser ignorado no in\'{\i}cio do c\'alculo.}


\subsection{O caso n\~ao adiab\'atico}

	Consideremos agora o caso de uma barra em que h\'a transmiss\~ao do calor. 
O estudo da evolu\c c\~ao do sistema obriga-nos a considerar, desde o in\'{\i}cio, 
n\~ao unicamente $X = X(x,t)$, mas tamb\'em $T = T(X,t)$, ou $T = T(x,t)$. \\ 

	J\'a n\~ao podemos usar as f\'ormulas (6), e temos de usar as f\'ormulas do 
tipo (4) que t\^em de ser conhecidas para resolver os problemas concretos. \\ 

	Temos de considerar, ainda, um fluxo de calor que, representaremos no referencial
pr\'oprio por $q_0$, e que ter\'a de ser convenientemente estimado.

	Localmente as componentes do tensor impuls\~ao energia, no referencial 
pr\'oprio $S^*$, devem ser: 

\be T^{\alpha^*\beta^*} = \left[ \begin{array}{cccc} 
p & 0 & 0 & \fr{q_0}{c} \\
0 & 0 & 0 & 0 \\
0 & 0 & 0 & 0 \\
\fr{q_0}{c} & 0 & 0 & \rho_0 c^2 \end{array} \right] \ee

	Num qualquer referencial de in\'ercia $S$, as componentes do mesmo tensor s\~ao: 

\be T^{\alpha \beta} = \left[ \begin{array}{cccc}
\fr{p + \beta^2 \rho_0 c^2 + \fr{2q_0}{c}\beta}{1 - \beta^2} & 0 & 0 &
\fr{\beta(p + \rho_0 c^2) + (1 + \beta^2)\fr{q_0}{c}}{1 - \beta^2} \\ 
0 & 0 & 0 & 0 \\ 
0 & 0 & 0 & 0 \\
\fr{\beta(p + \rho_0 c^2) + (1 + \beta^2)\fr{q_0}{c}}{1 - \beta^2} & 0 & 0 & 
\fr{\beta^2p + \rho_0 c^2 + \fr{2q_0}{c}\beta}{1 - \beta^2} 
\end{array} \right] \ee 

	As duas leis de conserva\c c\~ao traduzem-se, assim, pelas equa\c c\~oes: 
\be \begin{array}{c} \partial_x \left( \fr{p + \beta^2 \rho_0 c^2 + \fr{2q_0}{c}\beta}
{1 - \beta^2} \right) + 
\fr{1}{c} \partial_t \left( \fr{\beta(p + \rho_0 c^2) + (1 + \beta^2)\fr{q_0}{c}}
{1 - \beta^2} \right) = 0 \\  \\ 
 \partial_x \left(  \fr{\beta(p + \rho_0 c^2) + (1 + \beta^2)\fr{q_0}{c}}{1 - \beta^2} \right) + 
\fr{1}{c} \partial_t \left( \fr{\beta^2p + \rho_0 c^2 + \fr{2q_0}{c}\beta}{1 - \beta^2} \right) = 0
\end{array} \ee

	Note-se, desde j\'a, que estas equa\c c\~oes s\~ao invariantes na mudan\c ca 
das coordenadas do referencial $S$ para as coordenadas de um outro qualquer 
referencial de in\'ercia $S^{'} (x^{'},ct^{'})$, e que s\~ao equa\c c\~oes diferenciais 
de segunda ordem relativamente \`as duas "inc\'ognitas": $X = X(x,t)$ e $T = T(x,t)$. \\ 

	Para termos, no entanto, um sistema matem\'atico bem definido, que nos 
permita estudar a evolu\c c\~ao do sistema, precisamos de conhecer, al\'em das 
rela\c c\~oes "constitutivas" (4), uma conveniente rela\c c\~ao 
relativista que desempenhe o papel da hip\'otese de Fourier: 

\be q_0 = - K \frac{\partial T}{\partial X} ~~~~ \left( ou ~~~~ q_0 = - K 
\frac{\partial T}{\partial x} \right) \ee

	Tal como Fourier no s\'eculo XVIII, o que podemos aqui fazer \'e avan\c car com 
algumas hip\'oteses que nos pare\c cam simples, a ver o que d\~ao. Se com elas chegamos 
a equa\c c\~oes matem\'aticamente trat\'aveis cujos resultados sejam condizentes com 
as observa\c c\~oes (que, neste caso, poss\'{\i}velmente, s\'o poder\~ao ser feitas 
em Astrof\'{\i}sica), ficamos contentes.  

	Como hip\'oteses razo\'aveis, admitimos as duas hip\'oteses em que se desdobra a 
hip\'otese de Fourier, dado $x$ e $X$ serem agora distintos e, ainda, a hip\'otese 
dimensionalmente correcta que podemos construir com as grandezas dispon\'{\i}veis: 

\be q_0 = - K \left( \frac{\partial T}{\partial X} + R \fr{\gamma_0}{c^2}T \right) \ee
sendo $\gamma_0 $ a acelera\c c\~ao no referencial pr\'oprio.


\section{A tradicional aus\^encia da transmiss\~ao do calor na mec\^anica dos fluidos relativista}

	Em quase todos os tratados de Relatividade h\'a um cap\'{\i}tulo sobre a 
mec\^anica dos fluidos relativistas. Estes cap\'{\i}tulos assentam essencialmente, 
(embora alguns n\~ao explicitamente), na considera\c c\~ao de um tensor impuls\~ao 
energia $T^{\alpha \beta}$ que verifica a equa\c c\~ao que exprime as leis de 
conserva\c c\~ao fundamentais: 

\be \Delta_{\beta} T^{\alpha \beta} = 0 \ee 
e na aceita\c c\~ao de "rela\c c\~oes constitutivas" caracter\'{\i}sticas dos fluidos em estudo, 
essenciais para escrever as componentes de  $T^{\alpha \beta}$ no referencial pr\'oprio.

Mas como definir o referencial pr\'oprio de um fluido( no\c c\~ao obviamente 
local, dado que varia de ponto para ponto e de instante para instante)?

Para Lichnerowicz \cite{dois}, por exemplo, e pensamos que para a generalidade dos relativistas 
que se ocupam do assunto, o referencial pr\'oprio \'e o referencial $S^*$ em 
que s\~ao nulas as componentes $T^{i4}$ e $T^{4i}$ de $T^{\alpha \beta}$ 
(quando usamos a coordenada $(x^i , x^4 = ct)$). Por outras palavras, \'e o 
referencial em que as componentes de $T^{\alpha \beta}$ s\~ao do tipo (7) e 
n\~ao do tipo (12). \\ 

	Mas \'e esta defini\c c\~ao  fisicamente aceit\'avel? \\ 

	Consideremos o exemplo simples de um g\'as, formado por part\'{\i}culas 
todas iguais, contido num cilindro im\'ovel num referencial $S$, com as duas 
bases a temperaturas diferentes.

	Numa situa\c c\~ao estacion\'aria, o n\'umero de part\'{\i}culas que atravessa 
uma sec\c c\~ao {\it im\'ovel} do cilindro num e noutro sentido, \'e o mesmo. 
 No entanto, no caso das suas energias cin\'eticas m\'edias serem diferentes a 
sec\c c\~ao \'e atravessada por um fluxo de calor.

	Neste caso, no referencial $S$ do cilindro, temos $T^{14} = \fr{q_0}{c} \neq 0$. 
As componentes do tensor impuls\~ao energia s\~ao portanto do tipo (12) e n\~ao (7).
	Para termos um referencial $S^*$ que satisfa\c ca a condi\c c\~ao de Lichnerowicz, 
temos de procurar um referencial em que o cilindro se desloque com uma conveniente 
velocidade $v$ para ser $T^{i^* 4^*} = 0$. \\

	Qual \'e, ent\~ao, no caso de haver um fluxo de calor 
$q_0$ o bom referencial pr\'oprio? 

	O referencial $S$ do cilindro, em que o g\'as est\'a 
numa situa\c c\~ao est\'atica, ou o referencial $S^*$, que satisfaz \`a 
condi\c c\~ao de Lichnerowicz mas em que o g\'as est\'a em movimento? \\ 

	Parece-nos que, manifestamente, a defini\c c\~ao de Lichnerowicz s\'o 
pode ser aceite nos casos em que {\it n\~ao h\'a transmiss\~ao do calor}. Por 
outras palavras, uma mec\^anica dos fluidos em que seja aceite esta defini\c c\~ao 
\'e, \`a partida, uma mec\^anica dos fluidos adiab\'atica. Um contributo para 
ultrapassar esta situa\c c\~ao pode-se encontrar, por exemplo na refer\^encia 
\cite{tres}. \footnote[4]
{H\'a um resultado relativista relacionado com estas quest\~oes, devido a Planck \cite{quatro},
conhecido desde 1907 e que, no entanto, n\~ao \'e v\'alido nos casos n\~ao 
adiab\'aticos: a invari\^ancia relativista das press\~oes. No 
modelo de um g\'as contido num cilindro com igual n\'umero de part\'{\i}culas 
a passar de um lado para o outro com velocidades diferentes, um simples 
exerc\'{\i}cio escolar permite mostrar que, sendo $p_0$ a press\~ao sobre 
uma sec\c c\~ao im\'ovel e $q_0$ o fluxo de calor que a atravessa, a press\~ao 
sobre a mesma sec\c c\~ao num referencial em que ela esteja em movimento 
com a velocidade $v$ (com o sentido de $q$) \'e:
\bc $ p = p_0 + \fr{q_0 v}{c^2} $ \ec }



\end{document}